\documentclass{article} 
\usepackage{nips15submit_e,times}
\usepackage{hyperref}
\usepackage{url}
\usepackage[pdftex]{graphics}
\usepackage{amsmath,amssymb,rotating,multirow}
\usepackage{subfigure}
\usepackage[plain]{algorithm}
\usepackage{algpseudocode}
\usepackage{mathrsfs}

\title{Microclustering: When the Cluster Sizes Grow Sublinearly with
  the Size of the Data Set}

\author{
  Jeffrey W. Miller\thanks{Department of Statistical Science, Duke
    University} \And Brenda Betancourt\footnotemark[1] \And Abbas
  Zaidi\footnotemark[1] \And Hanna Wallach\thanks{Microsoft Research
    New York City and University of Massachusetts Amherst} \And
  Rebecca C. Steorts\footnotemark[1]
}

\nipsfinalcopy 

\newcommand{\comment}[1]{}
\newcommand{\bpi}{\boldsymbol{\pi}}
\newcommand{\g}{\,|\,}
\newcommand{\teq}{\!=\!}
\newcommand\iid{\mathrel{\stackrel{\makebox[0pt]{\mbox{\normalfont\tiny
          iid}}}{\sim}}}
\renewcommand{\liminf}{\operatornamewithlimits{\textrm{lim\,inf}}}

\begin{document}

\maketitle

\begin{abstract}
Most generative models for clustering implicitly assume that the
number of data points in each cluster grows linearly with the total
number of data points. Finite mixture models, Dirichlet process
mixture models, and Pitman--Yor process mixture models make this
assumption, as do all other infinitely exchangeable clustering
models. However, for some tasks, this assumption is undesirable. For
example, when performing entity resolution, the size of each cluster
is often unrelated to the size of the data set. Consequently, each
cluster contains a negligible fraction of the total number of data
points. Such tasks therefore require models that yield clusters whose
sizes grow sublinearly with the size of the data set. We address this
requirement by defining the \emph{microclustering property} and
introducing a new model that exhibits this property. We compare this
model to several commonly used clustering models by checking model fit
using real and simulated data sets.
\end{abstract}

\section{Introduction}

Many clustering tasks require models that assume cluster sizes grow
linearly with the size of the data set. These tasks include topic
modeling, inferring population structure, and discriminating among
cancer subtypes. Infinitely exchangeable clustering models, including
finite mixture models, Dirichlet process mixture models, and
Pitman--Yor process mixture models, all make this linear growth
assumption, and have seen numerous successes when used for these
tasks. For other clustering tasks, however, this assumption is
undesirable. One prominent example is entity resolution. Entity
resolution (including record linkage and deduplication) involves
identifying duplicate\footnote{In the entity resolution literature,
  the term ``duplicate records'' does not mean that the records are
  identical, but rather that they are corrupted, degraded, or
  otherwise noisy representations of the same entity.}  records in
noisy databases~\cite{christen12data,christen12survey}, traditionally
by directly linking records to one another. Unfortunately, this
approach is computationally infeasible for large data sets---a serious
limitation in ``the age of big
data''~\cite{christen12data,winkler06overview}. As a result,
researchers increasingly treat entity resolution as a
clustering task, where each entity is implicitly associated with one
or more records and the inference goal is to recover the latent
entities (clusters) that correspond to the observed records (data
points)~\cite{steorts??bayesian,steorts15entity,steorts14smered}. In
contrast to other clustering tasks, the number of data points in each
cluster remains small, even for large data sets. Tasks like this
therefore require models that yield clusters whose sizes grow
sublinearly with the total number of data points. To address this
requirement, we define the \emph{microclustering property} in
section~\ref{sec:microclustering} and, in section~\ref{sec:nbnb},
introduce a new model that exhibits this property. Finally, in
section~\ref{sec:experiments}, we compare this model to several
commonly used infinitely exchangeable clustering models.

\section{The Microclustering Property}
\label{sec:microclustering}

To cluster $N$ data points $x_1, \ldots, x_N$ using a partition-based
Bayesian clustering model, one first places a prior over partitions of
$[N] = \{ 1, \ldots, N \}$. Then, given a partition $C_N$ of $[N]$,
one models the data points in each part $c \in C_N$ as jointly
distributed according to some chosen form. Finally, one computes the
posterior distribution over partitions and, e.g., uses it to identify
probable partitions of $[N]$. Mixture models are a well-known type of
partition-based Bayesian clustering model, in which $C_N$ is
implicitly represented by a set of cluster assignments $z_1, \ldots,
z_N$. One regards these cluster assignments as the first $N$ elements
of an infinite sequence $z_1, z_2, \ldots$, drawn a priori from
\begin{equation}
  \bpi \sim H \quad \textrm{and} \quad
 z_1, z_2, \ldots \g \bpi \iid \bpi,
 \label{eqn:mixture}
\end{equation}
where $H$ is a prior over $\bpi$ and $\bpi$ is a vector of mixture
weights with $\sum_l \pi_l \teq 1$ and $\pi_l \geq 0$ for all
$l$. Commonly used mixture models include (a) finite mixtures where
the dimensionality of $\bpi$ is fixed and $H$ is usually a Dirichlet
distribution; (b) finite mixtures where the dimensionality of $\bpi$
is a random variable~\cite{richardson97bayesian,miller15mixture}; (c)
Dirichlet process (DP) mixtures where the dimensionality of $\bpi$ is
infinite~\cite{sethuraman94constructive}; and (d) Pitman--Yor process
(PYP) mixtures, which generalize DP
mixtures~\cite{ishwaran03generalized}.

Equation~\ref{eqn:mixture} implicitly defines a prior over partitions
of $\mathbb{N} = \{ 1, 2, \ldots \}$. Any random partition
$C_{\mathbb{N}}$ of $\mathbb{N}$ induces a sequence of random
partitions $(C_N : N=1, 2, \ldots)$, where $C_N$ is a partition of
$[N]$. Via the strong law of large numbers, the cluster sizes in any
such sequence obtained via equation~\ref{eqn:mixture} grow linearly
with $N$, since with probability one, for all $l$, $\frac{1}{N}
\sum_{n=1}^N I(z_n \teq l) \rightarrow \pi_l$ as $N \rightarrow
\infty$, where $I(\cdot)$ denotes the indicator
function. Unfortunately, this linear growth assumption is not
appropriate for entity resolution and other tasks that require
clusters whose sizes grow sublinearly with $N$.

To address this requirement, we therefore define the
\emph{microclustering property}: A sequence of random partitions $(C_N
: N=1, 2, \ldots)$ exhibits the microclustering property if $M_N$ is
$o_p(N)$, where $M_N$ is the size of the largest cluster in
$C_N$. Equivalently, $M_N \,/\, N \rightarrow 0$ in probability as $N
\rightarrow \infty$.



A clustering model exhibits the microclustering property if the
sequence of random partitions implied by that model satisfies the
above definition. No mixture model can exhibit the microclustering
property (unless its parameters are allowed to vary with $N$). In
fact, Kingman's paintbox
theorem~\cite{kingman78representation,aldous85exchangeability} implies
that any exchangeable partition of $\mathbb{N}$, such as a partition
obtained using equation~\ref{eqn:mixture}, is either equal to the
trivial partition in which each part contains one element or satisfies
$\liminf_{N \rightarrow \infty} M_N \,/\, N > 0$ with positive
probability. By Kolmogorov's extension theorem, a sequence of random
partitions $(C_N : N=1, 2, \ldots)$ corresponds to an exchangeable
random partition of $\mathbb{N}$ whenever (a) each $C_N$ is
exchangeable and (b) the sequence is consistent in
distribution---i.e., if $N' \!<\! N$, the distribution of $C_{N'}$
coincides with the marginal of $C_{N'}$ obtained using the
distribution of $C_N$. Therefore, to obtain a nontrivial model that
exhibits the microclustering property, one must sacrifice either (a)
or (b). Previous work~\cite{wallach10alternative} sacrificed (a);
here, we instead sacrifice (b).

\section{A Model for Microclustering}
\label{sec:nbnb}

In this section, we introduce a new model for microclustering. We
start by defining
\begin{equation}
  K \sim \textrm{NegBin}\,(a, q) \quad \textrm{and} \quad
 N_1, \ldots, N_K \g K \iid \textrm{NegBin}\,(r, p),
  \label{eqn:nbnb}
\end{equation}
for $a, r \,>\, 0$ and $q, p \in (0, 1)$. Note that $K$ and some of
$N_1, \ldots, N_K$ may be zero. We then define $N = \sum_{k=1}^K N_k$
and, given $N_1, \ldots, N_K$, generate a set of cluster assignments
$z_1, \ldots, z_N$ by drawing a vector uniformly at random from the
set of permutations of $(\underbrace{1,\ldots,1}_\text{$N_1$
  times},\underbrace{2,\ldots,2}_\text{$N_2$
  times},\ldots\ldots,\underbrace{K,\ldots,K}_\text{$N_K$ times})$.

The cluster assignments $z_1, \ldots, z_N$ induce a random partition
$C_N$ of $[N]$, where $N$ is itself a random variable---i.e., $C_N$ is
a random partition of a random number of elements. We call the
resulting marginal distribution of $C_N$ the NegBin--NegBin (NBNB)
model. If $\mathscr{C}_N$ denotes the set of all possible partitions
of $[N]$, then $\bigcup_{N=1}^{\infty} \mathscr{C}_N$ is the set of
all possible partitions of $[N]$ for $N \in \mathbb{N}$. In
appendix~\ref{sec:appendix_a}, we show that under the NBNB model, the
probability of any given $C_N \in \bigcup_{N=1}^{\infty}
\mathscr{C}_N$ is
\begin{equation}
  \label{eqn:prob_C_N_main_text}
  P(C_N) = \frac{p^N}{N!}\, a^{(|C_N|)}\,
  \frac{(1-q)^a \left(q\,(1-p)^r\right)^{|C_N|}}{\left(1 - q\,(1-p)^r\right)^{a + |C_N|}} \, \prod_{c \in
    C_N} r^{(|c|)},
  \end{equation}
where $x^{(m)} = x\,(x + 1) \ldots (x + m - 1)$ for $m \in \mathbb{N}$
and $x^{(0)} \teq 1$. We use $|\cdot|$ to denote the cardinality of a
set, so $|C_N|$ is the number of (nonempty) parts in $C_N$ and $|c|$
is the number of elements in part $c$. We also show that if we replace
the negative binomials in equation~\ref{eqn:nbnb} with Poissons, then
we obtain a limiting special case of the NBNB model. We call this
model the permuted Poisson sizes (PERPS) model. Under certain
conditions, the PERPS model is equivalent to the linkage structure
prior~\cite{steorts14smered,steorts??bayesian}.


In practice, $N$ is usually observed. Conditional on $N$,
the NBNB model implies that
\begin{equation}
  P(C_N \g N) \propto a^{(|C_N|)}\,\beta^{|C_N|} \, \prod_{c \in C_N}
  r^{(|c|)},
\end{equation}
where $\beta = \left(q\,(1 - p)^r\right) \,/\, \left( 1 -
q\,(1-p)^r\right)$. This equation leads to the following
``reseating algorithm''---much like the Chinese restaurant process
(CRP)---derived by sampling from $P(C_N \g N, C_N \!\setminus\! n)$,
where $C_N \!\setminus\!  n$ is the partition obtained by removing
element $n$ from $C_N$:
\begin{itemize}
\item for $n = 1,\ldots,N$, reassign element $n$ to
  \begin{itemize}
  \item an existing cluster $c \in C_N \!\setminus\! n$ with
    probability $\propto |c| + r$,
  \item a new cluster with probability $\propto
    (|C_N \!\setminus\! n|+ a)\,\beta r$.
  \end{itemize}
\end{itemize}
We can use this algorithm to draw samples from $P(C_N \g N)$, however,
unlike the CRP, it does not produce an exact sample if used to
incrementally construct a partition from the empty set. When the NBNB
model is used as the prior in a partition-based clustering
model---e.g., as an alternative to equation~\ref{eqn:mixture}---the
resulting Gibbs sampling algorithm for $C_N$ is similar to this
algorithm, but accompanied by appropriate likelihood
terms. Unfortunately, this algorithm is slow for large data sets. We
therefore propose a faster Gibbs sampling algorithm---the
\emph{chaperones algorithm}---in appendix~\ref{sec:appendix_b}.

In appendix~\ref{sec:appendix_c}, we present empirical evidence that
suggests that the sequence of partitions $(C_N : N = 1, 2, \ldots)$
implied by the NBNB model does indeed exhibit the microclustering
property.

\section{Experiments}
\label{sec:experiments}

In this section, we compare the NBNB and PERPS models to several
commonly used infinitely exchangeable clustering models: mixtures of
finite mixtures (MFM)~\cite{miller15mixture}, DP mixtures, and PYP
mixtures. We assess how well each model ``fits'' partitions typical of
those arising in entity resolution and other tasks involving clusters
whose sizes grow sublinearly with $N$. We use two observed
partitions---one simulated and one real. The simulated partition
contains 5,000 elements, divided into 4,500 clusters. Of these, 4,100
are singleton clusters, 300 are clusters of size two, and 100 are
clusters of size three. This partition represents data sets in which
duplication is relatively rare---roughly 91\% of the clusters are
singletons.  The real partition is derived from the Survey on
Household Income and Wealth (SHIW), conducted by the Bank of Italy
every two years. We use the 2008 and 2010 data from the Fruili region,
which consists of 789 records. Ground truth is available via unique
identifiers based upon social security numbers; roughly 74\% of the
clusters are singletons.

For each data set, we consider four statistics: the number of
singleton clusters, the maximum cluster size, the mean cluster size,
and the 90\% quantile of cluster sizes. We compare each statistic's
true value, obtained using the observed partition, to its distribution
under each of the models, obtained by generating 5,000 partitions
using the models' ``best'' parameter values. For simplicity and
interpretability, we define the best parameter values for each model
to be the values that maximize the probability of the observed
partition---i.e., the maximum likelihood estimate (MLE). The intuition
behind our approach is that if the observed value of a statistic is
not well-supported by a given model, even with the MLE parameter
values, then the model may not be appropriate for that type of data.

\begin{figure}[h]
  \begin{center}
    \includegraphics[width=1.0\textwidth]{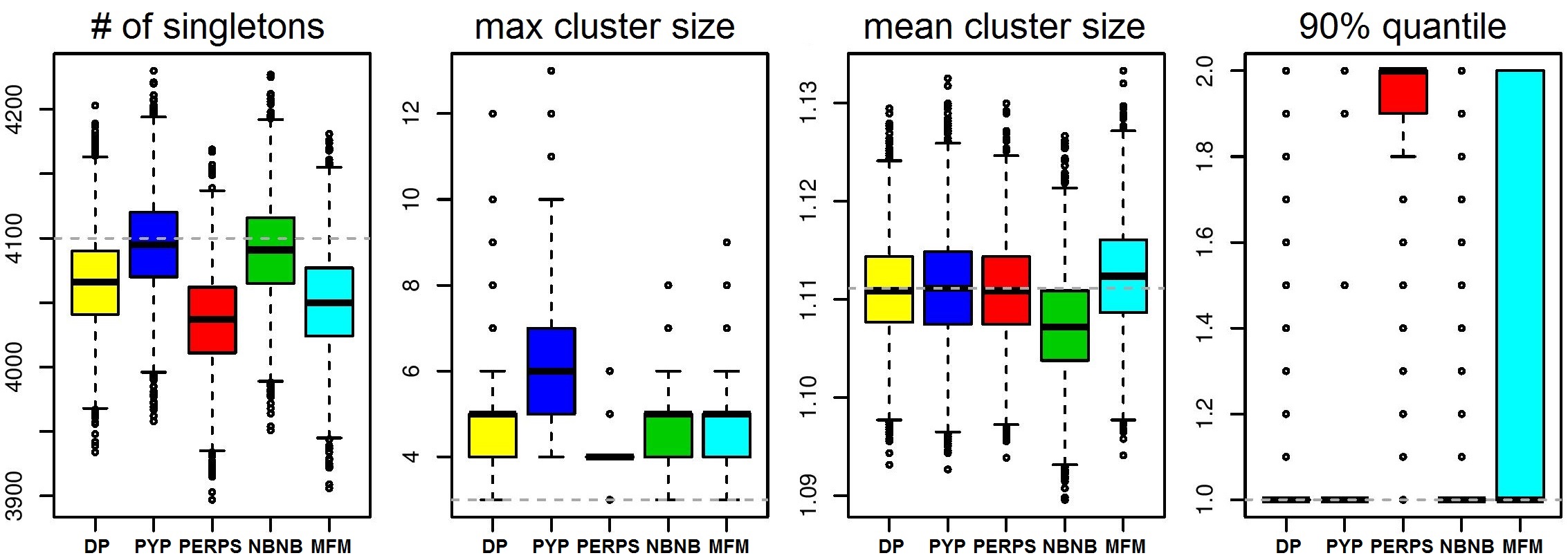}
    \caption{Results for the simulated partition. Each plot contains a
      boxplot depicting the distribution of a single statistic under
      each of the five models, obtained using the MLE parameter values
      (provided in appendix~\ref{sec:appendix_d}). The dashed
      horizontal line indicates the true value of the statistic.}
    \label{fig:simulated}
  \end{center}
\end{figure}

\begin{figure}[h]
  \begin{center}
    \includegraphics[width=1.0\textwidth]{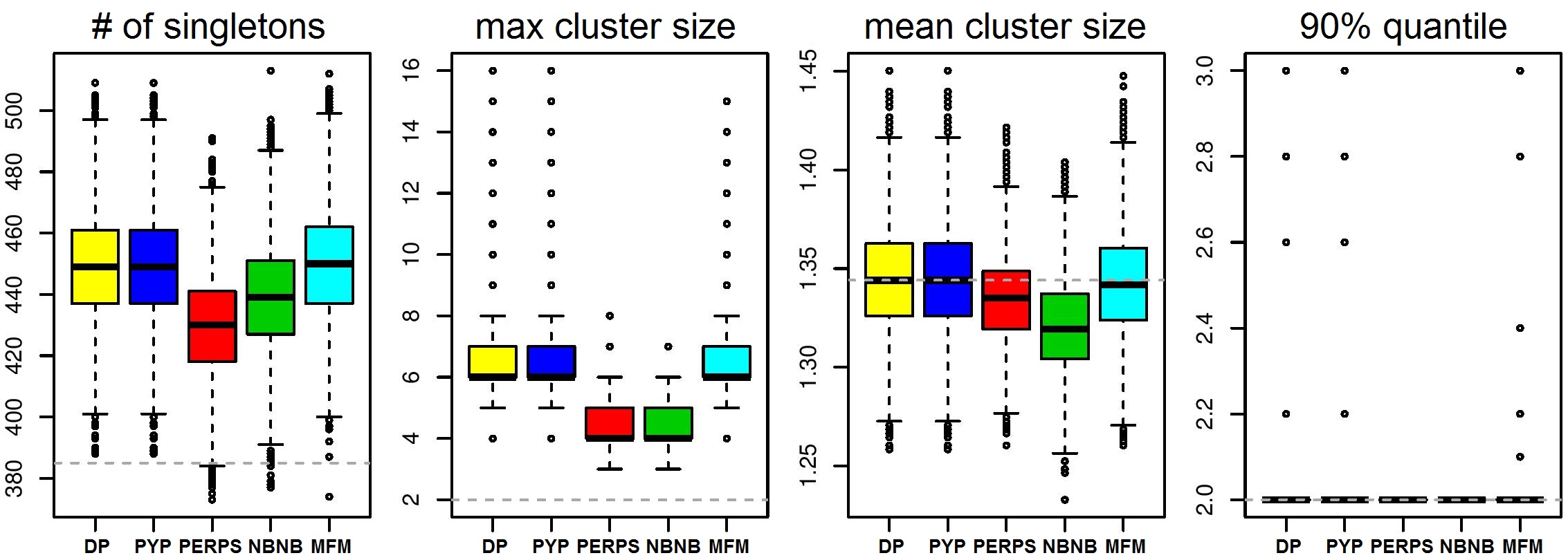}
    \caption{Results for the SHIW partition. This figure's
      interpretation is the same as that of
      figure~\ref{fig:simulated}.}
    \label{fig:real}
  \end{center}
\end{figure}

We provide plots summarizing our results in
figures~\ref{fig:simulated} and~\ref{fig:real}. The models are able to
capture the mean cluster size for each data set, although the NBNB
model's values are slightly low. For the SHIW partition, none of the
models do especially well at capturing the number of singleton
clusters or the maximum cluster size, although the NBNB and PERPS
models are the closest. For the simulated partition, neither the PYP
mixture model or the PERPS model are able to capture the maximum
cluster size. The PERPS model also does poorly at capturing the 90\%
quantile. Overall, the NBNB model appears to fit both data sets better
than the other models, though no one model is clearly superior to the
others. These results suggest that the NBNB model merits further
exploration as a prior for entity resolution and other tasks involving
clusters whose sizes grow sublinearly with $N$.

\section{Discussion}

Infinitely exchangeable clustering models assume that cluster sizes
grow linearly with the size of the data set. Although this assumption
is appropriate for some tasks, many other tasks, including entity
resolution, require clusters whose sizes instead grow sublinearly. The
microclustering property, introduced in
section~\ref{sec:microclustering}, provides a way to characterize
models that address this requirement. The NBNB model, introduced in
section~\ref{sec:nbnb}, exhibits this property. The figures in
section~\ref{sec:experiments} show that in some ways---specifically
the number of singleton clusters and the maximum cluster size---the
NBNB model can fit partitions typical of those arising in entity
resolution better than several commonly used mixture models. These
results suggest that the NBNB model merits further exploration as a
prior for tasks involving clusters whose sizes grow sublinearly with
the size of the data set.

\section*{Acknowledgments}

We would like to thank to Tamara Broderick, David Dunson, Merlise
Clyde, and Abel Rodriguez for conversations that helped form the ideas
contained in this paper. In particular, Tamara Broderick played a key
role in developing the idea of small clustering. JWM was supported in
part by NSF grant DMS-1045153 and NIH grant 5R01ES017436-05.  BB, AZ,
and RCS were supported in part by the John Templeton Foundation. RCS
was supported in part by NSF grant SES-1534412. HW was supported in
part by the UMass Amherst CIIR and by NSF grants IIS-1320219 and
SBE-0965436.

\bibliographystyle{unsrt} \bibliography{references}

\newpage

\appendix

\section{Derivation of $P(C_N)$ Under the NBNB and PERPS Models}
\label{sec:appendix_a}

In this appendix, we derive $P(C_N)$ where $C_N \in
\bigcup_{N=0}^{\infty} \mathscr{C}_N$. We start by noting that
\begin{equation}
  P(C_N) = \sum_{K=|C_N|}^{\infty} P(C_N \g K)\,P(K)
  \label{eqn:prob_C_N}
\end{equation}
and
\begin{equation}
  P(C_N \g K) = \sum_{z_1, \ldots, z_N \in [K]} \underbrace{P(C_N
    \g z_1, \ldots, z_N, K)}_{I(z_1, \ldots, z_N \Rightarrow C_N)}\,
  P(z_1, \ldots, z_N \g K)
\end{equation}
for any $K \geq |C_N|$. (The number of parts in $C_N$ may be less
than $K$ because some of $N_1, \ldots, N_K$ may be zero.) Since $N_1,
\ldots, N_K$ are completely determined by $K$ and $z_1, \ldots, z_N$,
\begin{align}
  P(z_1, \ldots, z_N \g K) &= P(z_1, \ldots, z_N \g N_1, \ldots,
  N_K, K)\, P(N_1, \ldots, N_K \g K)\\
  &= \frac{\prod_{k=1}^K N_k!}{N!} \prod_{k=1}^K P(N_k \g K) \\
  &= \frac{\prod_{k=1}^K N_k!}{N!} (1-p)^{Kr} \,p^N \prod_{k=1}^K
  \frac{r^{(N_k)}}{N_k!},
  \end{align}
where we have used $P(N_k \g K) = \frac{r^{(N_k)}}{N_k!} (1-p)^r\,
p^{N_k}$. Therefore,
\begin{align}
  P(C_N \g K) &= \sum_{z_1, \ldots, z_N \in [K]} I(z_1,\ldots, z_N
  \Rightarrow C_N) \, \frac{p^N}{N!} (1-p)^{Kr} \prod_{k=1}^K
  r^{(N_k)} \\
  &= \frac{p^N}{N!} (1-p)^{Kr} \left(\prod_{c\in C_N} r^{(|c|)}\right)
  \sum_{z_1, \ldots, z_N \in [K]} I(z_1, \ldots, z_N \Rightarrow
  C_N)\\
  &= \frac{p^N}{N!} (1-p)^{Kr} \left(\prod_{c\in C_N} r^{(|c|)}\right)
  \left( |C_N|!\right) {K \choose |C_N|}.
  \label{eqn:prob_C_N_g_K}
\end{align}
Substituting equation~\ref{eqn:prob_C_N_g_K} into
equation~\ref{eqn:prob_C_N} yields
\begin{equation}
  \label{eqn:prob_C_N_full}
  P(C_N) = \frac{p^N}{N!} \left(\prod_{c\in C_N} r^{(|c|)}\right)
  \sum_{K=|C_N|}^{\infty} (1-p)^{Kr}\left( |C_N|!\right) {K \choose |C_N|}\,
  P(K).
\end{equation}
Using $P(K) = \frac{a^{(K)}}{K!}  (1-q)^a q^k$, we know that
\begin{equation}
  \label{eqn:simplification}
  \sum_{K=|C_N|}^{\infty} (1-p)^{K r} \left(|C_N|!\right) {K \choose |C_N|}\,
  P(K)
  =
  \frac{(1-q)^a\left(q\,(1-p)^r\right)^{|C_N|}}{\left(1-q\,(1-p)^r\right)^{a+|C_N|}}
  \, a^{(|C_N|)}.
\end{equation}
Finally, substituting equation~\ref{eqn:simplification} into
equation~\ref{eqn:prob_C_N_full} yields
equation~\ref{eqn:prob_C_N_main_text} as desired.

The PERPS model is similar to the NBNB model
(equation~\ref{eqn:nbnb}), but with
\begin{equation}
  K \sim \textrm{Poisson}\,(\alpha) \quad \textrm{and} \quad
   N_1, \ldots, N_K \g K \iid \textrm{Poisson}\,(\lambda),
     \label{eqn:perps}
     \end{equation}
for $\alpha,\lambda>0$. If we let $r = \lambda\,/\,p$ and $a =
\alpha\,/\,q$, and then let $p,q \rightarrow 0$, then
equation~\ref{eqn:prob_C_N_main_text} converges to
\begin{equation}
  P_{\textrm{PERPS}}(C_N) = \frac{\lambda^N}{N!}\, \alpha^{|C_N|}\,
  e^{-\alpha}\, e^{-\lambda|C_N|}\, e^{\alpha e^{-\lambda}}.
  \label{eqn:prob_C_N_perps}
  \end{equation}
Equation~\ref{eqn:prob_C_N_perps} can also be derived directly by
following the approach used to derive
equation~\ref{eqn:prob_C_N_main_text}.

\section{The Chaperones Algorithm: Scalable Inference for Microclustering}
\label{sec:appendix_b}

For large data sets with many small clusters, standard Gibbs sampling
algorithms (such as the one outlined in section~\ref{sec:nbnb}) are
too slow to be used in practice. In this appendix, we therefore
propose a new sampling algorithm: the \emph{chaperones
  algorithm}. This algorithm is inspired by existing split--merge
Markov chain sampling
algorithms~\cite{jain04split--merge,steorts??bayesian,steorts14smered},
however, it is simpler, more efficient, and---most
importantly---likely exhibits better mixing properties when there are
many small clusters.

We start by letting $C_N$ denote a partition of $[N]$ and letting
$x_1, \ldots, x_N$ denote the $N$ observed data points. In the usual
incremental Gibbs sampling algorithm for nonparametric mixture models
(described in section~\ref{sec:nbnb}), each iteration involves
reassigning every element (data point) $n = 1, \ldots, N$ to either an
existing cluster or a new cluster by sampling from $P(C_N \g N, C_N
\setminus n, x_1, \ldots, x_N)$. When the number of clusters is large,
this step can be very inefficient because the probability that element
$n$ will be reassigned to a given cluster will, for most clusters, be
extremely small.

Our algorithm focuses on reassignments that have higher
probabilities. If we let $c_n \in C_N$ denote the cluster containing
element $n$, then each iteration of our algorithm consists of the
following steps:
\begin{enumerate}
\item Randomly choose two \emph{chaperones}, $i, j \in \{1, \ldots,
  N\}$ from a distribution $P(i, j \g x_1, \ldots, x_N)$ where the
  probability of $i$ and $j$ given $x_1, \ldots, x_N$ is greater
  than zero for all $i \neq j$. This distribution must be
  independent of the current state of the Markov chain $C_N$;
  however, crucially, it may depend on the observed data points
  $x_1, \ldots, x_N$.

\item Reassign each $n \in c_i \cup c_j$ by sampling from $P(C_N \g N,
  C_N \!\setminus\! n, c_i \cup c_j, x_1, \ldots, x_N)$.
\end{enumerate}

In step 2, we condition on the current partition of all elements
except $n$, as in the usual incremental Gibbs sampling algorithm, but
we also force the set of elements $c_i \cup c_j$ to remain
unchanged---i.e., $n$ must remain in the same cluster as at least one
of the chaperones. (If $n$ is a chaperone, then this requirement is
always satisfied.) In other words, we view the non-chaperone elements
in $c_i \cup c_j$ as ``children'' who must remain with a chaperone at
all times. Step 2 is almost identical to the restricted Gibbs moves
found in existing split--merge algorithms, except that the chaperones
$i$ and $j$ can also move clusters, provided they do not abandon any
of their children. Splits and merges can therefore occur during step
2: splits occur when one chaperone leaves to form its own cluster;
merges occur when one chaperone, belonging to a singleton cluster,
then joins the other chaperone's cluster.

This algorithm can be justified as follows: For any fixed pair of
chaperones $(i,j)$, step 2 is a sequence of Gibbs-type moves and
therefore has the correct stationary distribution. Randomly choosing
the chaperones in step 1 amounts to a random move, so, taken together,
steps 1 and 2 also have the correct stationary distribution (see,
e.g., \cite{tierney94markov}, sections 2.2 and 2.4). To guarantee
irreducibility, we start by assuming that $P(x_1, \ldots, x_N \g
C_N)\,P(C_N) > 0$ for any $C_N$ and by letting $C'_N$ denote the
partition of $N$ in which every element belongs to a singleton
cluster. Then, starting from any partition $C_N$, it is easy to check
that there is a positive probability of reaching $C'_N$ (and vice
versa) in finitely many iterations; this depends on the assumption
that $P(i, j \g x_1, \ldots, x_N) > 0$ for all $i \neq
j$. Aperiodicity is also easily verified since the probability of
staying in the same state is positive.

The main advantage of the chaperones algorithm is that it can exhibit
better mixing properties than existing sampling algorithms. If the
distribution $P(i, j \g x_1, \ldots, x_N)$ is designed so that $x_i$
and $x_j$ tend to be similar, then the algorithm will tend to consider
reassignments that have a relatively high probability. In addition,
the algorithm is easier to implement and more efficient than existing
split--merge algorithms because it uses Gibbs-type moves, rather than
Metropolis-within-Gibbs moves.

\section{NBNB and the Microclustering Property}
\label{sec:appendix_c}

\begin{figure}[h]
  \begin{center}
    \includegraphics[width=1.0\textwidth]{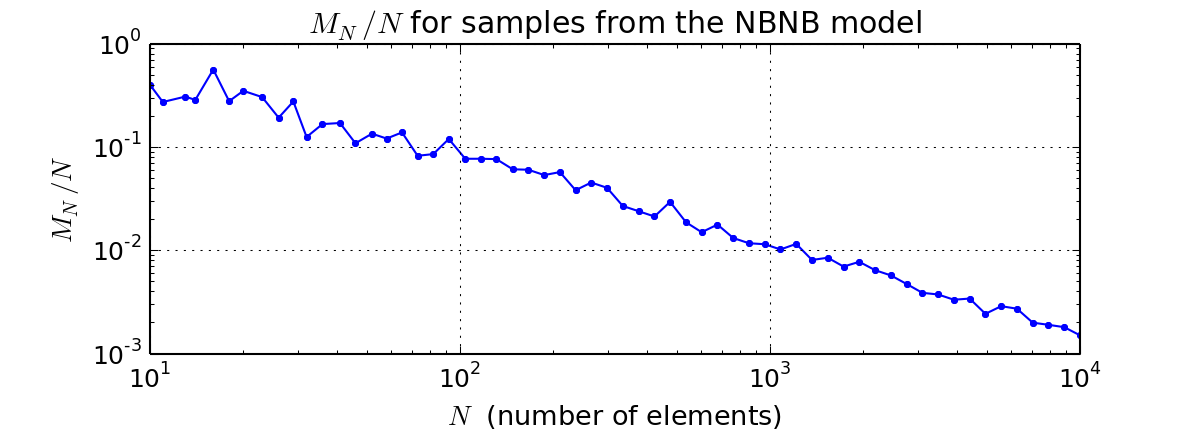}
    \caption{Empirical evidence suggesting that the NBNB model
      exhibits the microclustering property.}
    \label{fig:microclustering}
  \end{center}
\end{figure}

In this appendix, we present empirical evidence that suggests that the
sequence of partitions implied by the NBNB model exhibits the
microclustering property. Figure~\ref{fig:microclustering} shows $M_N
\,/\, N$ for samples of $M_N$ over a range of $N$ values from ten to
$10^4$. We obtained each sample of $M_N$ using the NBNB model with
$a=1$, $q=0.9$, and $r, p$ such that $\mathbb{E}(N_k \g K) = 3$ and
$\textrm{var}\,(N_k \g K)=3^2\,/\,2$. For each value of $N$, we
initialized the algorithm with the partition in which all elements are
in a single cluster. We then ran the reseating algorithm in
section~\ref{sec:nbnb} for 1,000 iterations to generate $C_N$ from
$P(C_N \g N)$, and set $M_N$ to the size of the largest cluster in
$C_N$. As $N \rightarrow \infty$, $M_N \,/\, N$ appears to converge to
zero in probability, suggesting that the model exhibits the
microclustering property.

\section{Experiments}
\label{sec:appendix_d}

For each data set, we calculated each model's MLE parameter values
using the Nelder--Mead algorithm. For the simulated partition, the MLE
values are $r = 2.955 \times 10^{-5}$, $p = 0.1875$, $a = 102.4$, and
$q = 0.9999$ for the NBNB model; $\lambda = 0.215$ for the PERPS
model; $\gamma = 0.279$ for the MFM model; $\theta = 21,\!719$ for the
DP model; and $\theta = 9,\!200$ and $\delta = 0.540$ for the PYP
model. For the SHIW partition, the MLE values are $r = 1,\!001$, $p =
6.212\times 10^{-4}$, $a = 100.6$, and $q = 0.9267$ for the NBNB
model; $\lambda = 0.624$ for the PERPS model; $\gamma = 1.056\times
10^{-2}$ for the MFM model; $\theta = 1,\!037$ for the DP model; and
$\theta = 1,\!037$ and $\delta = 0$ for the PYP model (making it
identical to the DP model). Note that for both data sets, the DP and
PYP concentration parameters are very large.

\comment{
\textbf{Number of singleton clusters:} Roughly 91\% of the clusters in
the simulated partition are singletons, compared to only 74\% in the
SHIW partition. For the simulated partition, the median number of
singleton clusters under the PYP model is closest to the true number;
for the SHIW partition, the PERPS model is closest. These results
suggest that the PYP model can (slightly) better fit the data when the
percentage of singleton clusters is higher; when the percentage of
singleton clusters is lower, the PERPS model is (slightly) better. For
both data sets, the NBNB model is a close second.

\textbf{Maximum cluster size:} For the SHIW partition, the NBNB and
PERPS models have median maximum cluster sizes that are closest to the
true maximum cluster size of two. For the simulated partition, the
PERPS model has the closest median maximum cluster size; however, the
true maximum cluster size is within the plot ``whiskers'' for the
NBNB, MFM, and DP models.

\textbf{Mean cluster size:}
\textbf{90\% quantile of cluster sizes:}
}

\end{document}